\documentclass[conference]{IEEEtran}
\IEEEoverridecommandlockouts
\usepackage{cite}
\usepackage{amsmath,amssymb,amsfonts}
\usepackage{algorithmic}
\usepackage{graphicx}
\usepackage{textcomp}
\usepackage{xcolor}
\usepackage{url}
\usepackage{booktabs}
\usepackage{listings}
\usepackage{multirow}
\usepackage{braket}
\usepackage{threeparttable}
\usepackage{ascmac}
\usepackage{subcaption}
\usepackage[a4paper, total={184mm,239mm}]{geometry}
\def\BibTeX{{\rm B\kern-.05em{\sc i\kern-.025em b}\kern-.08em
    T\kern-.1667em\lower.7ex\hbox{E}\kern-.125emX}}

\begin{document}

\title{Scoring-based Static Variable Ordering for Decision Diagram-based Quantum Circuit Simulation}

\author{\IEEEauthorblockN{
Yusuke Kimura\IEEEauthorrefmark{1}\IEEEauthorrefmark{4}, Masahiro Fujita\IEEEauthorrefmark{2}\IEEEauthorrefmark{3} and Robert Wille\IEEEauthorrefmark{4}\IEEEauthorrefmark{5}}
\IEEEauthorblockA{\textit{Fujitsu Limited, Japan}\IEEEauthorrefmark{1}}
\IEEEauthorblockA{\textit{The University of Tokyo, Japan\IEEEauthorrefmark{2}}
\IEEEauthorblockA{\textit{National Institute of Advanced Industrial Science and Technology (AIST), Tokyo, Japan}\IEEEauthorrefmark{3}}
\IEEEauthorblockA{\textit{Technical University of Munich, Munich, Germany}\IEEEauthorrefmark{4}}
\IEEEauthorblockA{\textit{Munich Quantum Software Company, Garching near Munich, Germany}\IEEEauthorrefmark{5}}
\\
yusuke-kimura@fujitsu.com\IEEEauthorrefmark{1},
fujita@ee.t.u-tokyo.ac.jp\IEEEauthorrefmark{2},
robert.wille@tum.de\IEEEauthorrefmark{4}}}

\maketitle

\begin{abstract}
Decision diagram (DD)-based quantum circuit simulators represent quantum states and gates using DDs, enabling memory-efficient and fast simulations for some quantum circuits like Shor. Although it is known that DD size and processing time vary depending on the variable order in classical circuits, there has not been much research on the variable order under quantum circuit simulation. One existing study pointed out that dynamic reordering worsens the simulation time and numerical accuracy, and there is no comprehensive research on static orders in the context of quantum circuit simulation. 
Therefore, this paper proposes a scoring-based heuristic method for determining a static variable order that enables efficient DD-based quantum circuit simulation. When applied to benchmark circuits, the default original variable orders resulted in slow simulations, whereas the proposed method achieved speedups of up to 150x. Furthermore, the proposed order completed the simulation of Shor's 1011 factorization in 5 hours on a single-core laptop, although it was not completed within two days previously.
\end{abstract}

\section{Introduction}
Recently, quantum computers have seen remarkable progress, with the development of various real machines~\cite{48651,ibmQuantumSystem}. However, accessing large-scale real machines is difficult for many researchers, and their cost is high. Furthermore, practical quantum error correction requires at least several thousand physical qubits per logical qubit~\cite{PRXQuantum.5.010337, PhysRevA.86.032324}, making it impossible to execute practical quantum algorithms such as Shor's algorithm on today's quantum computers. Current real quantum devices do not allow direct access to state vectors. Therefore, quantum circuit simulators running on current classical computers are used to develop quantum algorithms and quantum software.

The most common quantum circuit simulator is the explicit state vector-based one~\cite{aer,qulacs}. An $N$-qubit quantum state requires a complex vector of length $2^N$, and this type of simulator generally stores such a vector explicitly in memory. Despite its simple implementation and ability to be parallelized, it has the drawback that the required memory and computational complexity increase exponentially with the number of qubits.

To address the memory issue mentioned above, several alternative types of quantum circuit simulators have been proposed~\cite{Schollw_ck_2011,cuquantum, Bravyi_2019, ORUS2014117, Gray2021hyperoptimized, Fannes1992, Vidal_2003, Bridgeman_2017}. The decision diagram (DD)-based quantum circuit simulator~\cite{1623982,10.1109/TCAD.2015.2459034,8355954,8942057}, which is the focus of this paper, is one such simulator. It represents vectors and matrices using a graph structure called a decision diagram. As described later, it can reduce memory usage and computational complexity when subvectors have common or similar parts (such as the cases where a sub-vector is a scalar product of another), or when the number of non-zero values is small (sparse). In some quantum circuits, it can dramatically reduce simulation time compared to explicit state vector-based simulators~\cite{9308161,10821387}.

Decision diagrams have been used as a data structure for representing logical functions~\cite{1676819,10.1023/A:1008647823331}, and previous research revealed that their variable order significantly affects the size of diagrams and hence processing time in classical circuits~\cite{580029,122450, Qayyum2023, 10093493}. In DD for quantum computation, it was clarified that dynamic variable ordering is not useful due to the longer simulation time and numerical error~\cite{10.1007/978-3-031-09005-9_7}. 
Although some initial studies have been on static variable order~\cite{shen2022reordertrickdecisiondiagram,4341491}, no comprehensive research has covered various quantum circuits.
Therefore, this study aims to investigate the influence of static variable order on simulation time and to propose a method to determine an appropriate static variable order based on the features of a given quantum circuit.

The main contributions of this paper are as follows:
\begin{itemize}
\item To the best of our knowledge, this is the first paper to propose a method for determining the appropriate static variable order for DD-based quantum circuit simulation.
\item We found that the order of the variables significantly affects the simulation time of a DD-based quantum circuit simulator, and the proposed method achieves speedups of up to 150x.
\item The simulation of Shor's 1011 factorization finished within 5 hours on a single-core laptop, although it was not completed within two days using the default original order.
\end{itemize}

\section{Background \& Existing research}
\subsection{Quantum circuit simulation}\label{sec:quantum_basic}
This section explains a quantum circuit and its simulation. Operations performed on a quantum computer can be considered as a series of quantum gates applied to qubits, and those can be depicted as a quantum circuit. The quantum state of $N$ qubits is represented by a complex vector of length $2^N$. The value of each element corresponds to the state from $\ket{0\dots0}$ to $\ket{1\dots1}$, and the square of its absolute value represents the measurement probability when the qubits are observed. A quantum gate acting on $N$ qubits is represented by a unitary matrix of size $2^N \times 2^N$. We can calculate the product of the corresponding matrix and vector to apply a quantum gate to a particular quantum state. The newly obtained complex vector represents the quantum state after the gate operation.

\subsection{DD-based Quantum Circuit Simulation}
DDs have been used as data structures for representing logical functions~\cite{1676819,10.1023/A:1008647823331}, and they can also be used to represent vectors and matrices~\cite{1623982,10.1109/TCAD.2015.2459034,8355954,8942057}. DDSIM~\cite{8355954} and SliQSim~\cite{9586191} are famous DD-based quantum circuit simulator implementations. There is research using multi-process in an HPC environment~\cite{10646511}.

While there are various types of DDs, this paper focuses specifically on DD with edge values~\cite{8355954} (QDD). 
In QDD, the element value of the state vector can be obtained by tracing the edges according to the index value and calculating the product of the edge values. In the graphical representation, 0 in the index corresponds to moving left, and 1 in the index corresponds to moving right. An edge without an associated value is considered to be 1. The DD becomes canonical with a fixed index variable ordering by restricting how edge values are assigned.

\begin{figure}[tb]
  \centering
  \begin{minipage}{0.43\columnwidth}
      \includegraphics[width=\linewidth,trim={0 1cm 0 1cm}]{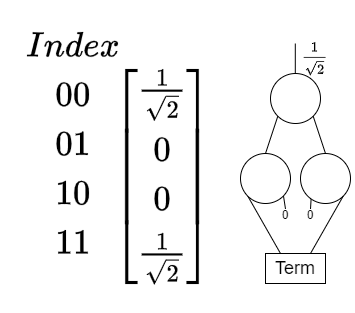}
      \caption{State vector representation in DD}
      \label{fig:dd:vector}
  \end{minipage}
  \begin{minipage}{0.55\columnwidth}
      \includegraphics[width=\linewidth,trim={0 1cm 0 1cm}]{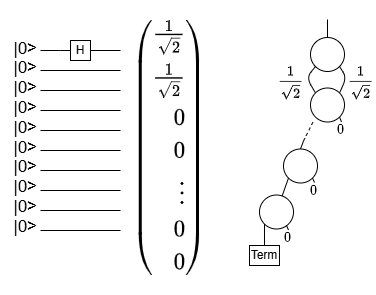}
      \caption{Example where DD reduces the memory amount significantly}
      \label{fig:dd:example}
  \end{minipage}
\end{figure}

\paragraph*{Example}
Consider the state vector in Fig.~\ref{fig:dd:vector}. To find the 4th value from DD, use 3 (counting from 0, binary representation: 11) as the index. In this case, follow the edges from the top in the order of right and right. The edge values are $(\frac{1}{\sqrt{2}},1,1)$, so the product is $\frac{1}{\sqrt{2}}$. If the index is 01, the product is 0 because an edge weight of 0 appears along the way. 

Fig.~\ref{fig:dd:example} illustrates the memory reduction effect of DD. It is a 10-qubit circuit with a Hadamard gate applied to the first qubit. The explicit state vector-based simulator must store $2^{10} = 1024$ complex numbers, while the DD needs only 10 nodes. In this way, DDs can reduce memory usage when the graph nodes are shared or when there are many common values or 0 (sparse) in the state vector.

\section{Motivation}
In decision diagram (DD)-based quantum circuit simulation, the simulation time can vary depending on the variable order.
The ordering can be categorized into two ways: dynamic or static.
Regarding DD-based quantum circuit simulation, one existing research~\cite{10.1007/978-3-031-09005-9_7} reported that dynamic reordering can increase simulation time and numerical errors. Therefore, this study focuses on the static variable order given before the simulation and does not address dynamic reordering.

There are several existing studies on static variable order. However, these studies only demonstrate that variable orders can affect simulation time and did not propose a new method to determine the order~\cite{4341491}, or focused on specific quantum algorithms~\cite{shen2022reordertrickdecisiondiagram}.

Specific examples of how static variable order affects DD are given below.
Fig.~\ref{fig:dd:qw} shows DDs for the 11-qubit "Quantum Walk (v-chain)" circuit included in MQT Bench~\cite{quetschlich2023mqtbench}.
The left uses the default original variable order written in the circuit, the middle is the reversed order, and the right is a randomly shuffled order. Although these represent the same state vector, the shapes and the number of nodes differ. 
A smaller number of nodes in DD requires less memory, and its simulation time may become shorter.
Therefore, using static variables that result in fewer nodes is desirable.

\begin{figure}[tb]
 \centering
 \begin{minipage}{0.30\linewidth}
  \centering
  \includegraphics[height=6cm]{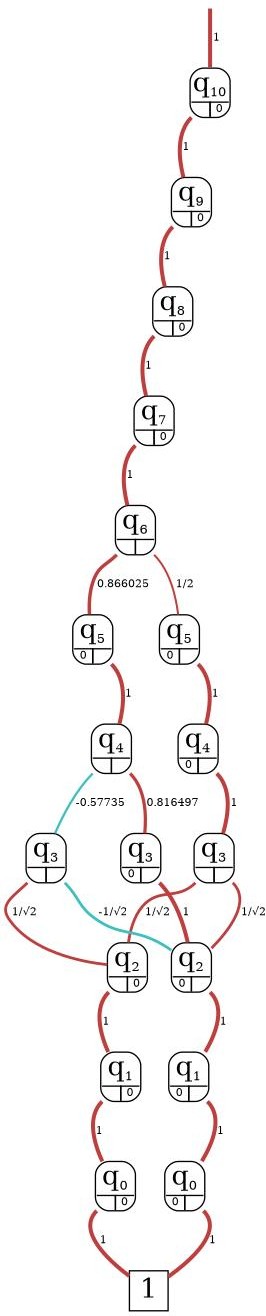}
 \end{minipage}
 \hspace{1mm}
 \begin{minipage}{0.30\linewidth}
  \centering
  \includegraphics[height=6cm]{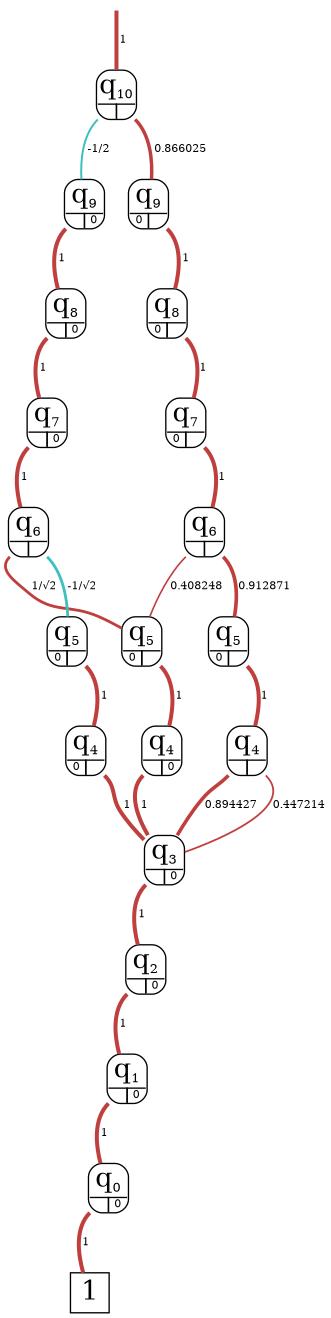}
 \end{minipage}
 \hspace{1mm}
 \begin{minipage}{0.30\linewidth}
  \centering
  \includegraphics[height=6cm]{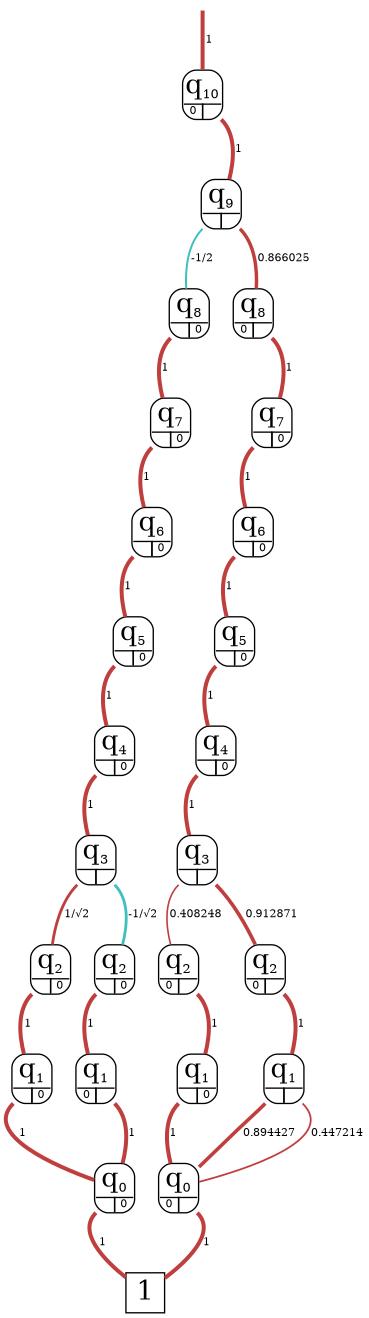}
 \end{minipage}
 \caption{Different orders for "Quantum Walk (v-chain)" 11-qubit circuit}
 \label{fig:dd:qw}
\end{figure}


\begin{figure}[tb]
    \centering
    \includegraphics[width=\linewidth,trim={0 0.5cm 0 1cm}]{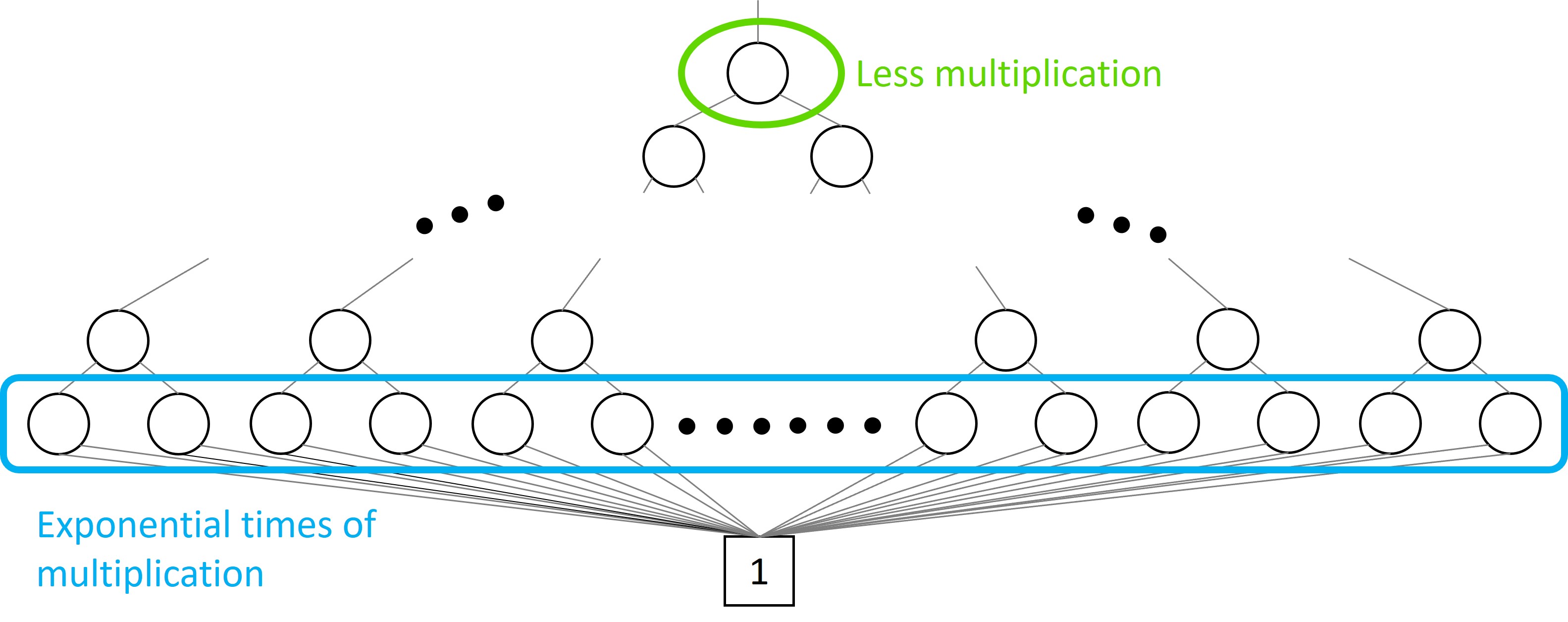}
    \caption{Qubit position and the computation complexity}
    \label{fig:multiply:amount}
\end{figure}

As described above, the static variable order affects the simulation time of DD-based quantum circuit simulators, but this has not been sufficiently studied.
Therefore, in this paper, we aim to propose a method for automatically determining static orders that minimize simulation time for given quantum circuits.

\section{Proposed method\\scoring-based heuristic static variable ordering}\label{sec:proposed}
This section proposes a scoring-based heuristic method for determining the static variable order that reduces DD-based quantum circuit simulation time.
The variable order is determined by assigning scores to each qubit in the following two steps, and qubits with higher scores are placed earlier in the variable order.
Code \ref{code:method} shows an overview of the algorithm.

\begin{enumerate}
\item Initial Scoring: Count the number of times each qubit acts as a control bit of a multi-bit gate, and predict the possibility of increasing the number of nodes in DD.
\item Score Manipulation: Count the number of parameterized rotation gates applied to each qubit, and adjust the score.
\end{enumerate}

\begin{figure}[tb]
\begin{lstlisting}[caption=Proposed method, label=code:method]
def sort_variables(circuit):
  nCtrl = {}      # Dict(qubit,count=0)
  nParamGate = {} # Dict(qubit,count=0)
  score = {}      # Dict(qubit,score=0)

  # Counting gates
  for gate in circuit:
      if isMultiGate(gate):
          for qubit in gate.ctrl_bits:
              nCtrl[qubit] += 1
      if hasParameter(gate):
          for qubit in gate.qubits:
              nParamGates[qubit] += 1

  # 1) Assign initial scores
  sorted_nCtrl = sort(nCtrl)
  tmp_score = 1 
  for qubit, count in sorted_nCtrl:
    score[qubit] = tmp_score
    tmp_score *= 2 # score becomes 1,2,4,8,...

  # 2) Manipulate scores
  for qubit, count in sorted_nCtrl:
    score[qubit] *= log(nParamGate[qubit])

  # Sort qubits (larger score is in front)
  sorted_score = sort(score, reversed=True)
  return sorted_score.keys() 
\end{lstlisting}
\end{figure}

It should be noted that the proposed method is heuristic and does not attempt to estimate the number of nodes or simulation time accurately. Instead, it is designed to predict the variable order in polynomial time with respect to the number of quantum gates.
In addition, the DD may contain floating-point errors and may not match the theoretically required number of nodes. Therefore, depending on the qubit order, the resulting state vector may slightly differ.

\subsection{Initial Scoring}
This section explains the background of using multi-bit gates in initial scoring.
In general, the number of nodes in DD increases when multi-bit gates are applied.
We decided to count the times each qubit becomes the control bit and assign a higher score to those with a higher number.
The qubits with a higher score are put in the upper part of the decision diagram.
This approach is expected to reduce the size of the DD.

Note that this scoring method is heuristic, since multi-bit gates may sometimes even decrease the number of nodes.
Also, the number of nodes may become $2^N$ for $N$-qubit circuits.
In such cases, the number of nodes is unchanged regardless of the qubit order, and the scoring in this section may not work well.

As explained in the previous section, DDSIM~\cite{8355954} also has a feature to decide the variable order by analyzing multi-bit gates. Although it requires a long time when the number of multi-bit gates is large, our proposed algorithm requires only $O(M)$ computation for $M$ gates. Therefore, the estimation accuracy can differ.

\subsection{Score adjustment}
This section explains the background of using parameterized rotation gates for score adjustment.
In DD-based quantum circuit simulators, the computation time tends to be long when parameterized rotation gates are applied.
For example, the following are unitary matrices representing an X gate without parameters and an RY gate, a parameterized rotation gate.
\begin{eqnarray*}
X=\begin{pmatrix}
   0 & 1 \\
   1 & 0
\end{pmatrix}, RY(\theta)=\begin{pmatrix}
   \cos{(\theta/2)} & -\sin{(\theta/2)} \\
   \sin{(\theta/2)} & \cos{(\theta/2)}
\end{pmatrix}
\end{eqnarray*}
For X gates, computation becomes simple using previous nodes because the values are only $0,1$.
On the other hand, for RY gates, depending on the value of $\theta$, it may be necessary to create new nodes, which can be time-consuming.

Therefore, we decided to count the times a parameterized rotation gate is applied to each qubit, and the score is multiplied by that count.
Through this score manipulation, qubits with many parameterized rotation gates can have earlier variable orders, which reduces the number of operations.
This score adjustment is useful when a DD has a tree structure. As shown in Fig.~\ref{fig:multiply:amount}, modifying only a few upper nodes is sufficient when applying a rotation gate to an earlier-ordered qubit. Conversely, computation is required for at most $O(2^N)$ nodes when applying a rotation gate to a later-ordered qubit. Thus, the adjustment in this sub-section is designed to reduce simulation time even when a DD is not tree and variable ordering cannot reduce the number of nodes.

The method is heuristic, so some adjustments are made.
For example, since not all rotation gates take a long time and to mitigate the impact of an extremely large number of rotation gates, the score is multiplied by a logarithm of the count, as in line 24.
Also, when the parameter is an integer multiple of $\frac{\pi}{2}$, we decided not to include them in the count.

\section{Experiment}\label{sec:experiment}
\subsection{Experimental Environment}
The experiments were conducted in a Linux environment (Ubuntu 24.04, Kernel: 5.15.167.4-WSL2) built on a Windows laptop (CPU: Core i7-1370P, Memory: 32GB). The memory allocated to the Linux environment was 16 GB. The DD-based simulator used in these experiments was MQT DDSIM (v1.24.1.dev57)~\cite{8355954}. All experiments were run three times, and the shortest simulation time was adopted, although the difference was insignificant. This is a similar setting to that in another simulator study~\cite{qulacs}.

TABLE~\ref{tab:orders} lists the variable orders implemented in this experiment. In addition to the proposed method, we implemented four variable orders based on the previous study of classical circuits~\cite{10093493}. Notably, not all variable orders were unique, as some produced the same orders. Also, the "DDSIM" variable order sometimes did not work for circuits with many gates, so a timeout of 10 minutes was set for calculating each variable order.

\begin{table}[t]
  \centering
\begin{threeparttable}[tb]
  \caption{List of variable orders in the experiments}
  \label{tab:orders}
  \begin{tabular}{cp{6cm}}\toprule
  Name & Explanation\\\midrule
  Original & The original default order written in the circuit\\
  Reversed & The reversed orders of the above original order\\
  nGates & The more gates are applied to a qubit, the earlier its order becomes.\\
  DDSIM\tnote{a}& Heuristic method implemented in MQT-DDSIM\cite{8355954}. It utilizes the control-target relationships of multiple CX gates\cite{4341491}.\\
  Proposed & Our proposed method\\\bottomrule
  \end{tabular}
  \begin{tablenotes}
    \item[a] \url{https://github.com/munich-quantum-toolkit/ddsim/pull/407}
  \end{tablenotes}
\end{threeparttable}
\end{table}

\subsection{Experiment.1: Analysis on Benchmark circuits}\label{sec:experiment1}
\subsubsection{Benchmark Circuits}
The quantum circuits used in this experiment were from MQT Bench~\cite{quetschlich2023mqtbench}. This benchmark suite provides circuits with 2 to 130 qubits for each quantum algorithm. In this experiment, we used the target-independent level circuits for Qiskit. The circuit size was chosen so that the simulation time would be, at most, about 10 minutes.

\subsubsection{Results}
The results are shown in TABLE~\ref{tab:result:mqt-bench}.
The left part shows the names of the circuits and the number of qubits and gates. For explanatory purposes, the circuits are divided into three groups: A, B, and C.
The runtimes for each variable order are shown on the right. The numbers of nodes are shown only for the original and proposed orders.
The fastest order did not change depending on the number of qubits among the same quantum algorithm.
Therefore, the table includes only the most significant number of qubit results for each quantum algorithm.

\begin{figure*}[tb]
  \centering
  \begin{minipage}{0.49\linewidth}
    \includegraphics[width=\linewidth,trim={2cm 3cm 0cm 3cm}]{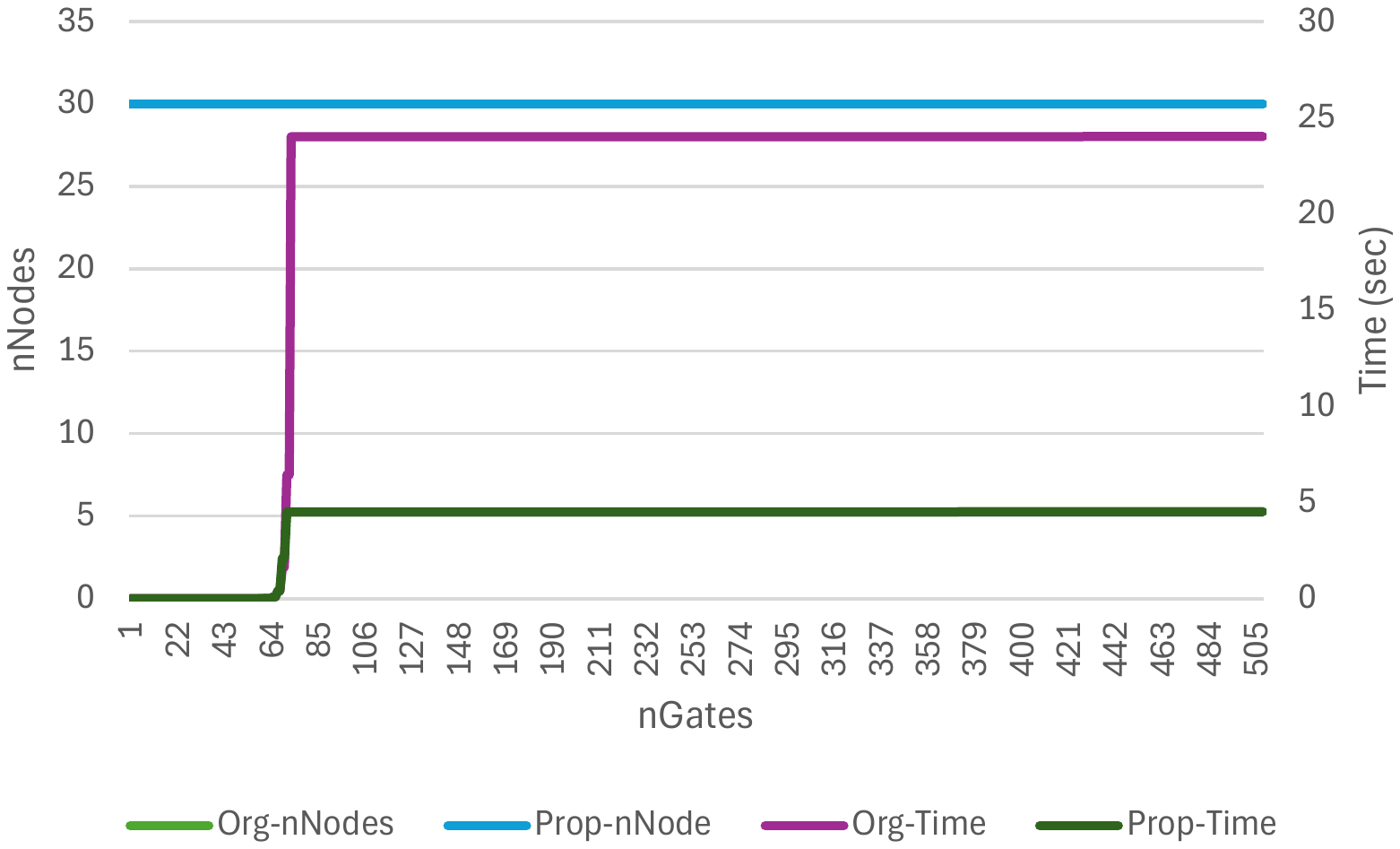}
    \subcaption{QPE exact circuit}
  \end{minipage}
  \begin{minipage}{0.49\linewidth}
    \includegraphics[width=\linewidth,trim={2cm 3cm 2cm 3cm}]{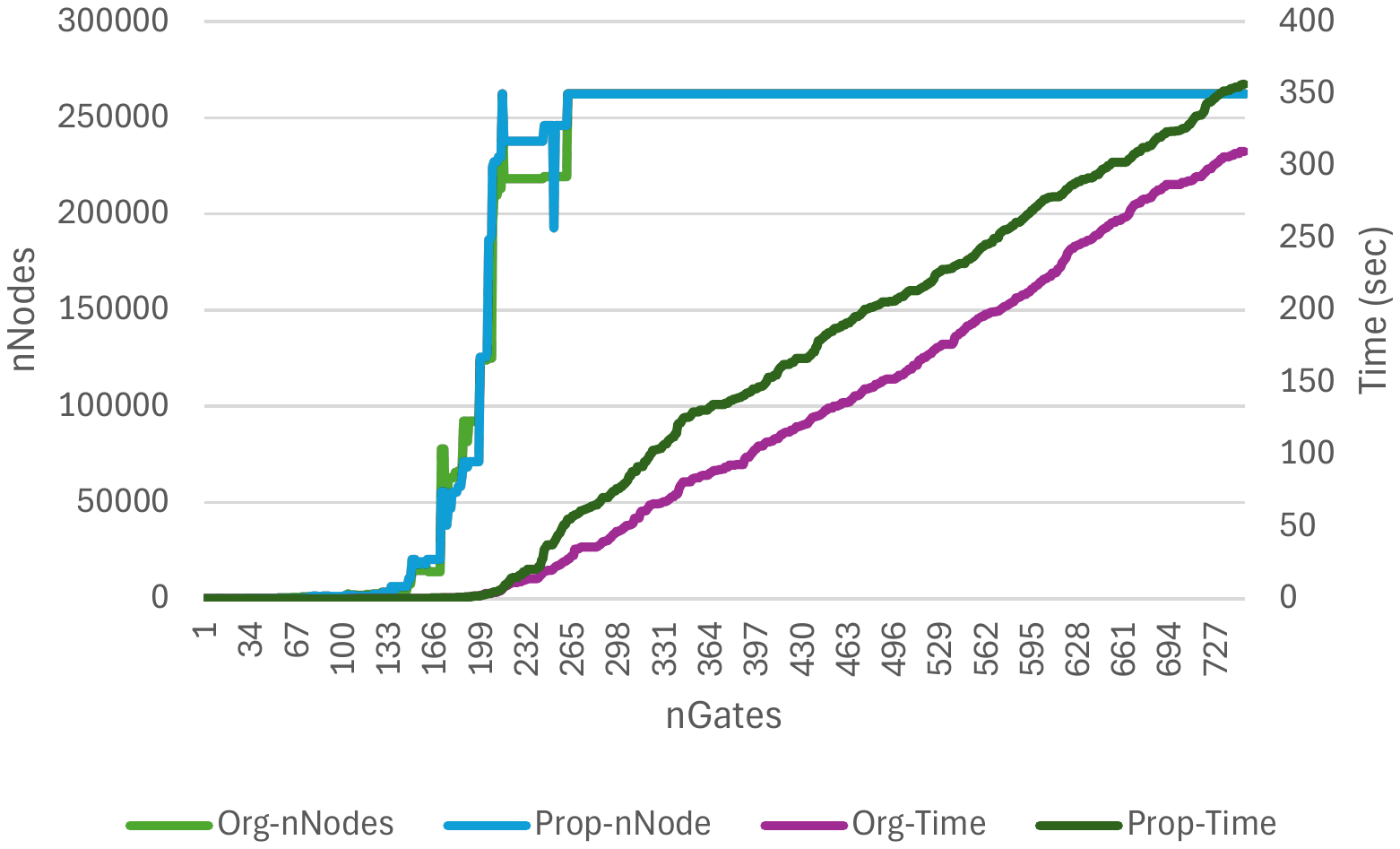}
    \subcaption{Random circuit}
  \end{minipage}
  
  \begin{minipage}{0.49\linewidth}
    \includegraphics[width=\linewidth,trim={2cm 3cm 0cm 2cm}]{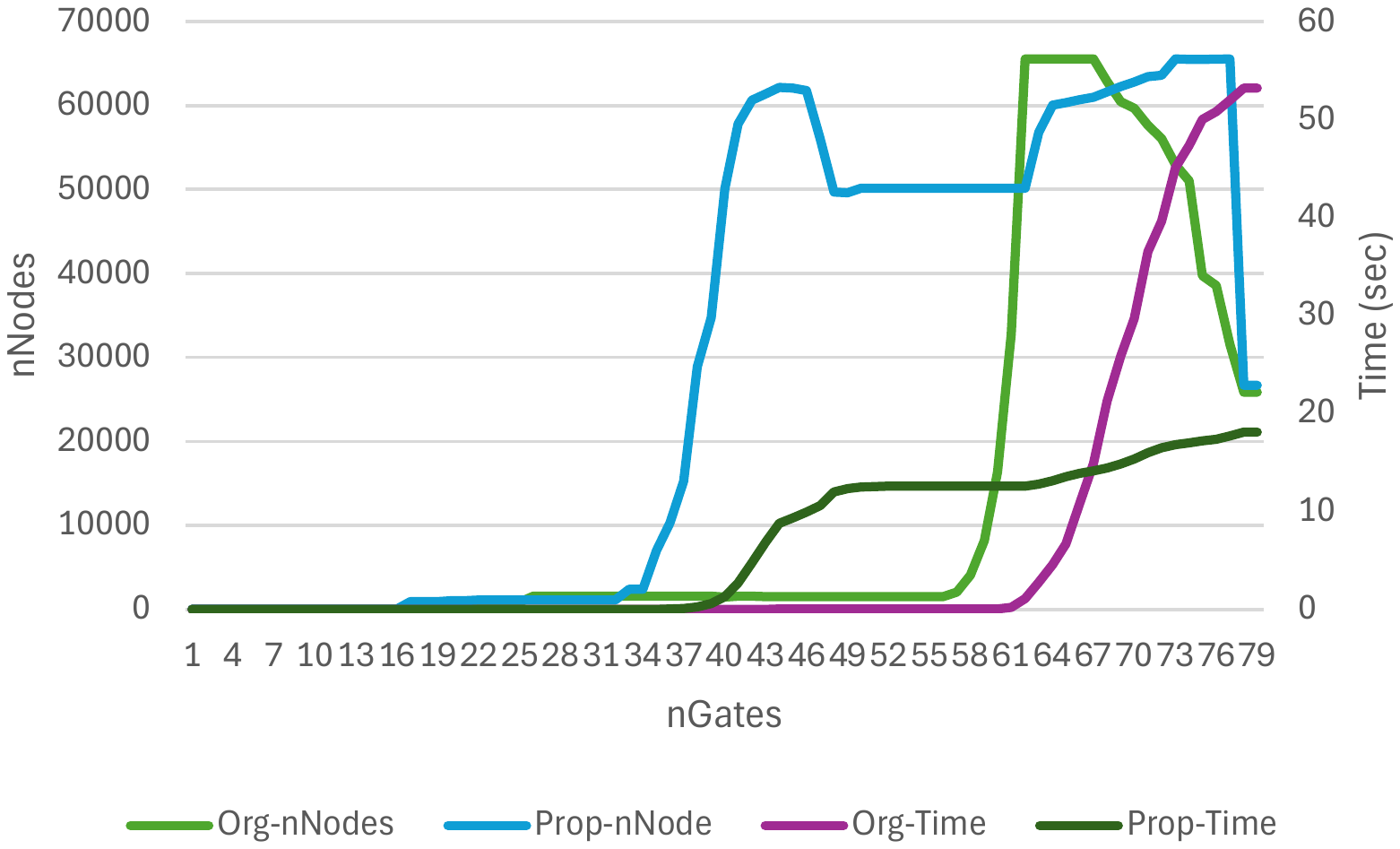}
    \subcaption{VQE circuit}
  \end{minipage}
  \begin{minipage}{0.49\linewidth}
    \includegraphics[width=\linewidth,trim={1cm 3cm 2cm 2cm}]{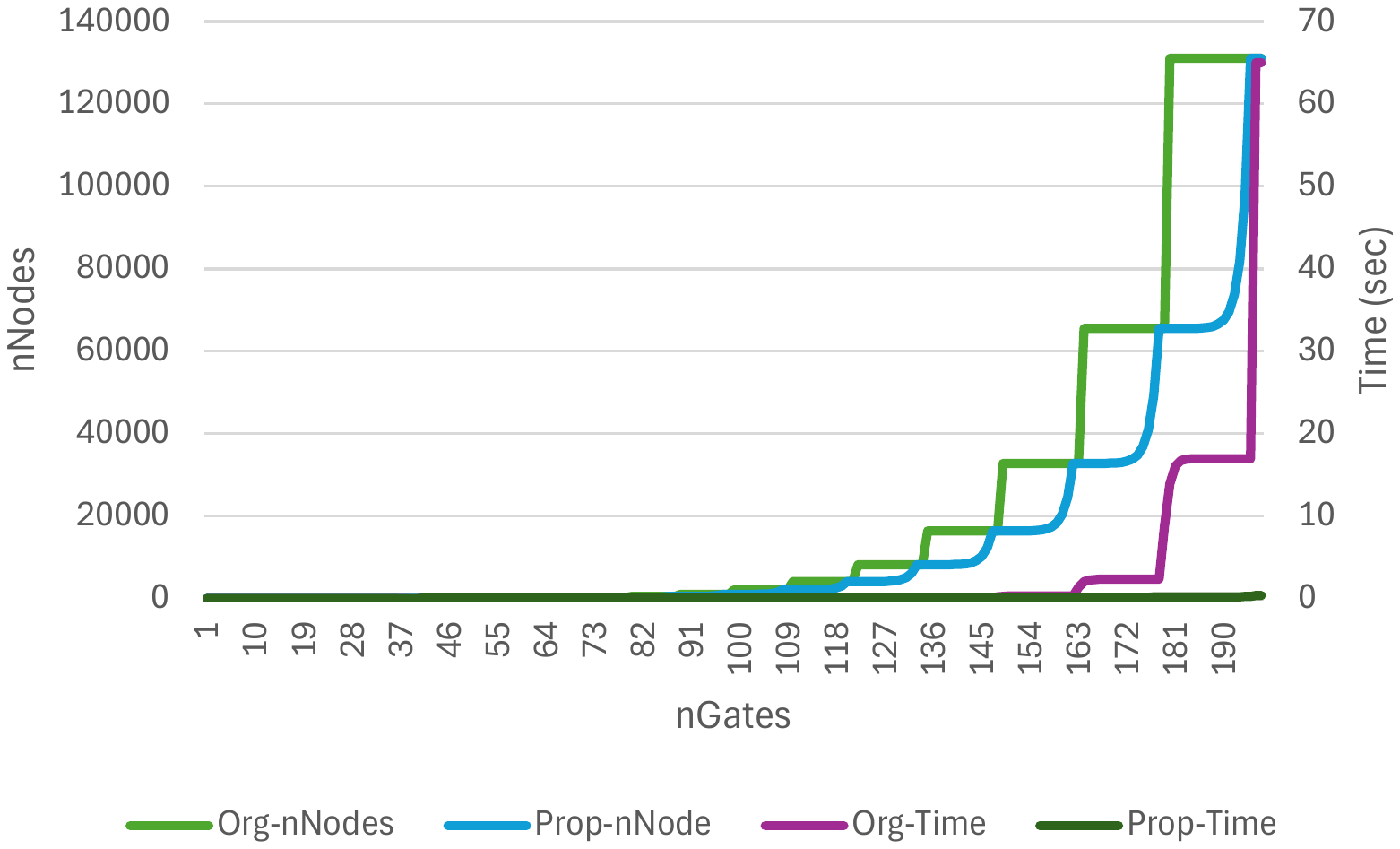}
    \subcaption{QPE inexact circuit}
  \end{minipage}
  \caption{The number of graph nodes and elapsed time during simulation (Horizontal: the number of gates processed)}
  \label{fig:ex1:stats}
\end{figure*}

Fig.~\ref{fig:ex1:stats} shows the number of graph nodes and elapsed time during the simulation. The horizontal axis represents the number of simulated gates, the left vertical axis represents the elapsed time, and the right vertical axis represents the number of graph nodes. 
Only the essential results for the four quantum circuits (QPE exact/inexact, random, VQE) are shown due to space constraints.
The simulation times in these graphs differ from those in TABLE~\ref{tab:result:mqt-bench} because additional calculations were performed to count the number of graph nodes at every gate.

\begin{table*}[t]
\centering
\begin{threeparttable}[t]
\caption{Benchmark circuits experiments}
\label{tab:result:mqt-bench}
\begin{tabular}{llrr|rrrrr|rrrr}   \toprule
\multirow{2}{*}{} & \multirow{2}{*}{Name} & \multirow{2}{*}{\#Qubits} & \multirow{2}{*}{\#Gates} & \multicolumn{5}{|c}{Simulation Time (sec)~\tnote{a}}            & \multicolumn{2}{|c}{\#Nodes (Original)}& \multicolumn{2}{c}{\#Nodes (Proposed)}\\\cmidrule(l){5-13} 
                  &                       &                         &  & Original & Reversed & nGates & DDSIM & Proposed & Final\tnote{b} & Max\tnote{c} & Final\tnote{b} & Max\tnote{c}  \\\midrule
 \multirow{9}{*}{A}
 & Grover’s (no ancilla) & 11 & 77,161 & \underline{0.63} & \underline{1.25} & \underline{0.86} & \underline{0.62} & \underline{0.77} & 28 & 56 & 29 & 81 \\
 & Q. Walk (no ancilla) & 13 & 98,325 & \underline{0.45} & \underline{0.49} & \underline{0.49} & \underline{1.19} & \underline{0.56} & 13 & 57 & 13 & 87\\
 & Grover’s (v-chain) & 17 & 52,011 & \underline{0.68} & \underline{1.4} & \underline{0.69} & \underline{0.6} & \underline{0.65} & 27 & 106 & 27 & 105\\
 & QPE exact & 30 & 507 & 27 & 32 & 14 & 27 & \underline{6} & 30 & 30 & 30 & 30\\
 & Q. Walk (v-chain) & 31 & 1,453& \underline{0.06} & \underline{0.078} & \underline{0.087} & \underline{0.067} & \underline{0.06} & 48 & 92 & 48 & 92 \\
 & Deutsch-Jozsa & 84\tnote{d} & 250 & \underline{0.043} & \underline{0.032} & \underline{0.043} & \underline{0.028} & \underline{0.043} & 84 & 84 & 84 & 84 \\
 & GHZ State & 130 & 130 & \underline{0.052} & \underline{0.036} & \underline{0.036} & \underline{0.052} & \underline{0.052} & 259 & 259 & 259 & 259 \\
 & QFT & 84\tnote{d} & 3,612 & \underline{0.064} & \underline{0.055} & \underline{0.064} & \underline{0.064} & \underline{0.045} & 84 & 84 & 84 & 84 \\
 & W-State & 130 & 517 & \underline{0.061} & \underline{0.050} & \underline{0.051} & \underline{0.050} & \underline{0.047} & 259 & 259 & 259 & 386 \\\midrule
  \multirow{10}{*}{B}
 & QAOA & 16 & 80 &\underline{0.11} & \underline{0.28} & \underline{0.11} & \underline{0.11} & \underline{0.26} & 11,116 & 13,940 & 19,479 & 25,602\\
 & Portfolio QAOA & 17 & 476 & \underline{25} & 28 & \underline{25} & \underline{25} & \underline{25} & 131,071 & 131,071 & 131,071 & 131,071\\
 & Amplitude Estimation & 18 & 239 & 45 & 369 & \underline{35} & 369 & \underline{38} & 262,143 & 262,143 & 262,143 & 262,143\\
 & Efficient SU2 ansatz & 18 & 531 &\underline{496} & 1,099 & \underline{496} & 1,099 & \underline{496} & 262,143 & 262,143 & 262,143& 262,143\\
 & Portfolio VQE & 18 & 531 &\underline{41} & 46 & \underline{41} & \underline{41} & \underline{41} & 262,143 & 262,143 & 262,143 & 262,143\\
 & QNN & 18 & 1,007 &\underline{24} & \underline{26} & 29 & 29 & \underline{24} & 262,143 & 262,143& 262,143 & 262,143\\
 & Random Circuit & 18 & 747 &\underline{279} & 374 & \underline{258} & T/O (10m)\tnote{e} & 370 & 262,143 & 262,143& 262,143 & 262,143\\
 & Real Amplitudes & 18 & 531 &\underline{482} & \underline{453} & \underline{482} & \underline{482} & \underline{482} & 262,143 & 262,143& 262,143 & 262,143\\
 & Two Local ansatz & 18 & 531 &\underline{439} & 491 & \underline{439} & 491 & \underline{439} & 262,143 & 262,143& 262,143 & 262,143\\
 & Graph State & 48 & 96 &\underline{0.48} & 24 & \underline{0.48} & \underline{0.48} & \underline{0.48} & 146,685 & 146,685 & 146,685 & 146,685 \\\midrule
\multirow{3}{*}{C}
 & VQE & 16 & 78 & 55 & \underline{3} & 74 & \underline{3} & 19 & 25,838 & 65,535 & 26,681 & 65,506\\
 & QFT Entangled & 18 & 198 & 553 & 535 & \underline{234} & 553 & \underline{245} & 262,143 & 262,143 & 262,143 & 262,143 \\
 & QPE inexact & 18 & 196 & 59 & \underline{0.61} & 59 & 59 & \underline{0.39} & 131,072 & 131,072 & 131,072 & 131,072\\\bottomrule
\end{tabular}
\begin{tablenotes}
\item[a] The runtime has underlines when it is the best or close. (Difference is 10\% / 1 sec or less than the best)
\item[b] The number of DD nodes representing the final state vector.
\item[c] The maximum number of nodes in DD during simulation.
\item[d] DDSIM had an error with an 85-qubit circuit or more.
\item[e] DDSIM heuristic took more than 10 minutes.
\end{tablenotes}
\end{threeparttable}
\end{table*}

\paragraph{Group A (nNodes=$O(N)$)}

The quantum circuits classified into this group are characterized by having few parameterized rotation gates and multi-qubit gates. In such cases, the state vector can be represented with $O(N)$ nodes for $N$ qubits. DD-based simulators can simulate them quickly regardless of the number of qubits.

Nine quantum algorithms were classified into this group. The number of nodes varied for some variable orders. However, since the simulation finishes very quickly regardless of the variable order, it is generally unnecessary to use a specific order. 
Since these circuits have almost no entanglement, the simulation time remains short even with a larger number of qubits, and this trend is likely to continue.

The QPE exact circuit is an exception, where the proposed method significantly reduced simulation time.
Fig.~\ref{fig:ex1:stats}(a) shows the number of nodes and time per gate, and it indicates that the specific part in the circuit consumes a long time.
We have manually confirmed that those parts include multi-bit gates, which are applied to earlier and later qubits in the variable order.
Even if the number of nodes in the state vector is small, such long-range multi-bit gates sometimes increase the simulation time.
Again, this case is rare because there are no branches in the DD, meaning no entanglement.

\begin{itembox}[l]{Analysis on Group A}
The simulation time of Group A circuits is usually very short, so there is no need to consider the variable order.
\end{itembox}

\paragraph{Group B (nNodes $\simeq O(2^N)$)}

Many parameterized rotations and multi-qubit gates characterize this group's circuits. In such cases, DD becomes complex and requires $2^N$ or a similar number of nodes for $N$ qubits. 
Ten quantum algorithms were classified into this group. Since the simulation time tends to be long, most experiments were conducted with up to 18-qubit circuits. 

There was no difference in the number of nodes regardless of the qubit orders, but our proposed method achieved up to 50x faster simulation than the worst orders. Please note that the original default order was also fast enough.

The random circuit was an exception; the original order was faster than the proposed one. The statistic per gate is shown in Fig.~\ref{fig:ex1:stats}(b). The number of nodes is increasing rapidly in the proposed method, which resulted in a longer simulation time. It was also observed that, after the number of nodes reached $2^N$, the simulation speed remained similar regardless of the variable order. Since the proposed method is a heuristic approach, it does not always find the optimal variable order.

\begin{itembox}[l]{Analysis on Group B}
The simulation time differs by the variable order up to 50 times.
The runtime of the proposed order is the shortest or close to the best result. Usually, the original order is fast enough for this group.
\end{itembox}

\paragraph{Group C (QPE, QFT, VQE)}\label{sec:experiment1:c}
Three quantum algorithms were classified into this group. 
The circuits in this group are similar to those in Group B, with many parameterized rotations and multi-qubit gates.

It is important to note that the default original order takes longer simulation time than the proposed one. In contrast to Groups A and B, it is clarified that variable ordering is crucial for DD-based simulators.

For the QPE inexact circuit, the simulation time with the proposed order was 150x shorter than that with the original order.
The background of this speedup can be the number of computations.
The number of computations increases when many gates are applied to the later qubits, especially when DD is a tree. 
As discussed in Fig.~\ref{fig:multiply:amount} of Sec.~\ref{sec:proposed}, our proposed method considers the number of gates applied to each qubit.
According to Fig.~\ref{fig:ex1:stats}(d), some gates took long times, and it was confirmed by manual inspection that those gates were applied to the later qubits. 

The initial state creation was the only difference between the QPE exact (Group A) and QPE inexact circuits (Group C).
The QFT (Group A) and QFT entangled (Group C) have the same feature.
The two paired circuits were similar, so the proposed variable orders became the same. However, the number of DD nodes varies significantly depending on the initial state, indicating that the DD size estimation from a given quantum circuit is difficult.

The optimal variable order could not be obtained for VQE circuits. 
As shown in Fig.~\ref{fig:ex1:stats}(c), the number of nodes exhibits significant fluctuations during VQE simulation, making it difficult to predict the variable order for a shorter simulation time.

\begin{itembox}[l]{Analysis on Group C}
It was found that the simulation time becomes long with the original orders, indicating the importance of static ordering. The proposed method achieved faster simulation than the original orders.
\end{itembox}

\subsection{Experiment. 2: Shor's Algorithm}\label{sec:experiment2}

We also used Shor's algorithm circuits. 
Shor is an important FTQC algorithm with exponential speedup, and it is also one of the algorithms that a DD-based quantum circuit simulator can run faster than other types of simulators~\cite{9308161,10821387}. Therefore, speeding up the DD simulation of Shor's algorithm is important.

The Shor's circuits were from the existing study~\cite{10646511}, using $4N+2$ qubits for an $N$-bit integer.
While more advanced methods have been proposed, such as factoring using only $2N+2$ qubits~\cite{10.5555/2011665.2011669}, this experiment used the $4N+2$-qubit circuit for ease of design.

\begin{table*}[t]
\centering
\begin{threeparttable}[t]
\caption{Shor's algorithm experiment}
\label{tab:result:shor}
\begin{tabular}{rrr|rrrrr|rr}   \toprule

\multirow{2}{*}{\begin{tabular}{c}Composite\\Number\end{tabular}} & \multirow{2}{*}{\#Qubits} & \multirow{2}{*}{\#Gates} & \multicolumn{5}{|c}{Simulation Time (sec)~\tnote{a}}            & \multicolumn{2}{|c}{\# Nodes in DD\tnote{b}}\\\cmidrule(l){4-10} 
                       &                         & & Original & Reversed & nGates & DDSIM & Proposed       & Original & Proposed \\\midrule 
21 & 24 & 25,330 & \underline{0.13} & \underline{0.22} & \underline{0.33} & \underline{0.22} & \underline{0.16} & 3,094 & 3,091 \\
57 & 26 & 51,123 & \underline{0.63} & \underline{1.5} & 8.3 & \underline{1.5} & \underline{0.92} & 36,885 & 25,622 \\
123 & 30 & 118,342 & \underline{1.8} & 4.3 & 16 & 4.3 & \underline{2.7} & 82,000 & 65,589 \\
253 & 34 & 204,637 & 1,116 & 247 & 1,474 & 247 & \underline{117} & 3,604,439 & 711,390 \\
511 & 38 & 400,123  & 596 & \underline{32} & 431 & \underline{32} & 57 & 786,511 & 786,503 \\
1,011 & 42 & 534,524 & $>$2 days & $>$2 days & $>$2 days & $>$2 days & \underline{18,144} & N/A & 7,830,983 \\\bottomrule
\end{tabular}
\begin{tablenotes}
\item[a] The runtime has underlines when it was the fastest or close. (Difference was 10\% / 1 sec or less than the best)
\item[b] The number of DD nodes representing the final state vector.
\end{tablenotes}
\end{threeparttable}
\end{table*}

The experimental results are shown in TABLE~\ref{tab:result:shor}. The first column is the numbers to be factored. Although the original variable orders were faster for smaller circuits up to 123, the time difference was slight. On the other hand, for larger circuits of 253 or more, the proposed variable orders achieved a speedup of about 10 times. For the factorization of 1011, the simulation with the other orders did not finish even after two days, while the proposed variable order was completed in about 5 hours. 
Furthermore, in all experiments in this subsection, the number of DD nodes in the proposed method was fewer than the original.
As shown above, it is clarified that our proposed method can be applied to large-scale practical circuits.

\section{Conclusions}
In this study, we investigated the effect of static variable order on the simulation time of a decision diagram(DD)-based quantum circuit simulator. We proposed a method for determining the variable order to shorten the simulation time.
Specifically, we determine the variable order by assigning scores to each qubit, considering the number of times it becomes a control bit in multi-bit gates and the number of parameterized rotation gates.

Benchmark experiments demonstrated that the proposed method can significantly reduce simulation time compared to the original variable order.
Furthermore, experiments using Shor's algorithm demonstrated that the proposed method can simulate large circuits that could not be simulated previously within a reasonable time.
The findings of this study are expected to contribute to the research and development of quantum algorithms using DD-based quantum circuit simulators.

Possible future works are as follows. Although we only focused on static variable ordering in this research, the combination with dynamic reordering might speed up the simulator.
Moreover, the experiments in this paper were conducted without considering noise. We would also like to investigate the effect of variable order with noisy DD-based simulation~\cite{9794651}.
Finally, we would like to mention the numerical errors.
As with other methods, DDs use floating-point numbers, and the values of edges that should be the same sometimes differ by a small amount.
It makes it difficult to predict the optimal variable order. Simulation algorithms that are less susceptible to such errors are also a future work.

\newpage
\bibliographystyle{IEEEtran}
\bibliography{bib}

\end{document}